\documentclass{elsart}

 \usepackage{graphics}
 \usepackage{graphicx}
 \usepackage{epsfig}

\usepackage{amssymb}

\newcounter{bla}

\begin{document}

\begin{frontmatter}



\title{HepML, an XML-based format for describing simulated data in high energy physics}

\author[jinr]{S.~Belov} 
\author[sinp]{L.~Dudko} 
\author[jinr]{D.~Kekelidze} 
\author[sinp,oxford]{A.~Sherstnev\corauthref{corts}} 

\address[jinr]{Joint Institute for Nuclear Research, Dubna, Moscow region, Russia} 
\address[sinp]{Scobeltsyn Institute of Nuclear Physics, Moscow State University, Moscow, Russia}
\address[oxford]{R.~Peierls Centre for Theoretical Physics, University of Oxford, Oxford, UK} 

\corauth[corts]{Corresponding author. \textit{Email address}: \texttt{a.sherstnev1@physics.ox.ac.uk}.}

\begin{abstract}
In this paper we describe a HepML format and a corresponding C++ library 
developed for keeping complete description of parton level events 
in a unified and flexible form. HepML tags contain enough information 
to understand what kind of physics the simulated events describe and how 
the events have been prepared. A HepML block can be included into 
event files in the LHEF format. The structure of the HepML block is 
described by means of several XML Schemas. The Schemas define necessary 
information for the HepML block and how this information should be located 
within the block. The library {\it libhepml} is a C++ library intended for 
parsing and serialization of HepML tags, and representing the HepML block 
in computer memory. The library is an API for external software. For example, 
Matrix Element Monte Carlo event generators can use the library for preparing 
and writing a header of a LHEF file in the form of HepML tags. In turn, 
Showering and Hadronization event generators can parse the HepML header and 
get the information in the form of C++ classes. {\it libhepml} can be used 
in C++, C, and Fortran programs. All necessary parts of HepML have been 
prepared and we present the project to the HEP community. 
\end{abstract}


\end{frontmatter}

\section{Introduction}
\label{intro}
In the last 10-15 years Monte-Carlo simulation programs have become one of
the main research tools both in phenomenological studies and experimental
analyses in high energy physics (HEP). This resulted in a burst of new
programs available for researchers. Since several such programs should be
usually interfaced to each other in practical calculations, importance of 
the interfacing is rising.
Generally, programs can be interfaced either in computer memory, via shared
libraries or unified executables, or externally, via data files. The latter
approach is more flexible and simpler in realization, although it can be
less reliable, since data files can be corrupted, lost, etc. Currently
intermediate data files, as a means of interfacing programs and a method
of data storage, are widely spread in HEP. In this paper we propose a new
markup language for a unified description of Monte-Carlo simulated events at
the parton level. This format, called High energy physics Markup Language
(HepML), is an extension for the Les Houches agreements and an attempt to
overcome several limitations of the Les Houches event format 
(LHEF)~\cite{Alwall:2006yp}.

Lets describe the problem we are going to tackle in this paper in a stricter
manner. In order to simulate collisions of two particles at an accelerator
in a physical model realistically we have to pass through several steps of
simulation (see more details in~\cite{Dobbs:2004qw}). At first, we generate
so called ``partonic events'' for a production process of one or several
particles we are interested in. The events are points in the phase space,
distributed according to the process matrix element squared. Examples of
programs, which can prepare such events, are ALPGEN~\cite{Mangano:2002ea},
CompHEP~\cite{Boos:2004kh}, MadGraph~\cite{Maltoni:2002qb}, 
HELAC~\cite{Kanaki:2000ey}, Whizard/OMega~\cite{Ohl:2000pr}, 
AMEGIC++~\cite{Krauss:2001iv}, Comix~\cite{Gleisberg:2008fv}, and 
Grace~\cite{Ishikawa:1991qb}. However,
these events are not what we can observe in particle detectors, since most
the final particles in the events are not real physical degrees of
freedom. The final partons should form hadrons, which can decay; the final
leptons can irradiate photons. This phase of simulation is called showering
(since the most important effect added here is the QCD showers), hadronization,
and decays. Currently, the most important players at this stage of simulation 
are Pythia~\cite{Sjostrand:2008vc}, Herwig++~\cite{Bahr:2008pv}, and
Sherpa~\cite{Gleisberg:2008ta}. After applying these effects we receive
events with observable particles in the finale state. However, these events
are not what experimentalists in HEP are interested in, since detector
effects must be added into the events in order to receive realistic
simulated detector output. Since the effects are detector dependent every
HEP experiment develops its own simulation software, in most cases not
publicly available. Different agreements and file formats exist
for transferring of simulated data through the whole simulation chain. As we
state above the HepML format is thought to be a part of the LHEF format,
which is currently a standard format for storing partonic event files. In
our opinion, the main limitation of LHEF is rigidity of the format structure.
Certainly it is not a problem to develop a simple and compact record for
several phase space points - momenta of all particles in an event and
several additional numbers, which characterize the event. But LHEF does not
have any means to keep an information on physical model parameters, applied
cuts, and other highly important information. The main obstacle for that is
internal heterogeneity of the information. Thus we are trying to propose a
rather simple format, which can include blocks of data with various
structure. Parsers of the format should be relatively simple and be able
to parse the HepML block if it contains superfluous information. This format
should be also based on some standard programming tools.

\begin{figure}[hbtp]
\begin{center}
\includegraphics[width=130mm,height=61mm]{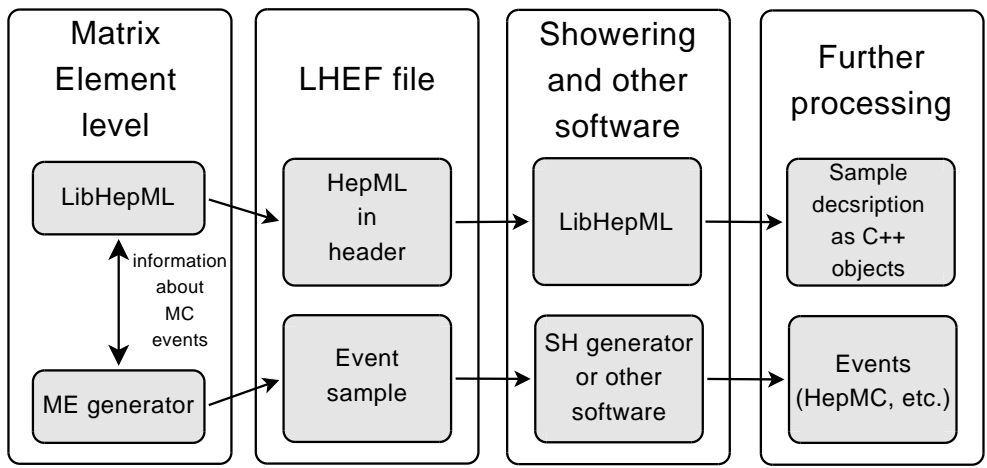}
\end{center}
\caption{
Place of HepML in the Monte Carlo simulation chain in HEP.
}
\label{fig:hepml_mc_simulation}
\end{figure}

Therefore our main goal is to overcome rigidity of the LHEF format by means
of some standard tools. To achieve this we chose the ideology of markup
languages~\cite{wikipedia_markup}. Markup languages, strictly
speaking procedural markup languages, are perfectly suitable for this goal,
since the languages are used for describing structure of complex data.
XML~\cite{Bray:2008tb}, as a standard instrument for developing markup
languages, is the most appropriate base for such format. Thus, HepML is a
markup language describing structure of data within the LHEF format.
Employing the standard software technologies allows users and developers to
re-use a lot of reliable and well-designed software, developed in the
industry of software development. Fig. ~\ref{fig:hepml_mc_simulation}
represents a place of the HepML language and libraries in the full simulation
chain in HEP.

XML has been applied in many programming projects in HEP, mainly in experimental
HEP. In particular, the CMS experiment stores detector geometry data in
XML~\cite{Arce:2003xs}, ATLAS uses XML tools for management of analysis
software~\cite{Obreshkov:2006qm}, Monte-Carlo simulation program 
Geant4~\cite{Allison:2006bu} stores detector geometry by means of a special
markup language GDML~\cite{Chytracek:1999bu}. XML-based formats also prove
their usefulness in many other scientific areas. We mention several 
successful examples only: 
\begin{itemize}
\item 
  Chemical Markup Language~\cite{Murray:1999pm} formalizes the structure of
  information about molecules, chemical reactions, analytical data, etc.
  It is extensively used in computational chemistry, quantum chemistry,
  material sciences. A lot of chemistry oriented software support the
  standard.
\item 
  MathML~\cite{Ausbrooks:2003ra} aims at providing ability to use mathematical
  notations in HTML documents. The format is supported in all major Web
  browsers and in an enormous number of applications, such as programs for
  distant education, computer algebra systems, formula editors, etc.
\item 
  CellML~\cite{Lloyd:2004rg} was originally created for describing
  mathematical models in cardiac research. Now it is used as a description 
  format for computational models, in many other areas of biology and other 
  sciences, such as the computer science.
\end{itemize}


Traditional event file formats, such as standard file headers in
CompHEP~\cite{Boos:2004kh} and MadGraph\cite{Maltoni:2002qb}, a format of
output files of showering/decay Monte Carlo generators
HepMC~\cite{Dobbs:2001ck}, have a fixed structure. This approach has both
benefits and drawbacks. Generally, a format of this type can look simpler
and more human-readable, than XML code, in ordinary text editors. It also
requires less programming efforts. However, any small correction of an
output/input file format of a program forces to modify both the program
itself and all programming clients of the program. As we mentioned earlier,
current simulation chain in HEP is rather long and quite complicated. So,
constant modifications in some parts of the chain cause work in other parts.
More stable file formats, such as HepML, will prevent this.

The next section conveys general conceptions of HepML. HepML XML Schemas are
described in the section~\ref{schema}. Since the format itself is a markup
language only, a software library for interpretation of the language is
necessary. The section~\ref{code} describes a programming API for this
format. It is a C++ library, called {\it libhepml}.
The section~\ref{howto} explains how to use the library. To date we have two
software projects which already adapt the HepML format and/or the library.
The section~\ref{application} outlines these projects and a role of HepML
in them. We summarize conclusions and future plans in the
section~\ref{conclusions}.

\section{General description}
A HepML block in a LHEF event files is an XML document with a pre-defined
structure. i.e. it is a piece of ordinary text marked up with tags. The text
has to contain an information enough to understand what kind of physics the events
describe and how the events have been prepared. The only place in a LHEF file,
which can accommodate the HepML block, is the tag \verb|<header><\header>|,
because the LHEF format does not define contents of the tag. Since a LHEF file
is \textit{not} an XML document, the file with a HepML block can not be an
XML document too. We intentionally do not propose to store event records in
the XML style, since this will add repeated tags in the records and makes the
event file larger. Since a typical event file can be huge with hundreds of
thousands of event records, these new tags in the event records would be too
expensive for manipulating with the files. On the other hand, extracting
contents of the tag \verb|<header><\header>| is quite simple programming task.

Tags in an HepML block are arranged in a determined structure. Several XML
Schemas~\cite{Thompson:2004tb} define the pre-defined structure. Tags
used in the block can be either standard ones, i.e defined in the XML Schemas,
or user-defined tags. The latter tags must obey XML rules only. 
The HepML XML Schemas define data types
and relative disposition of all tags in an XML document. The main principle
we keep in the setup can be formulated in the following manner: ``do not
remove or change the standard tags, add your own tags''. This means we assume
users will use existing tags from the HepML Schemas, will not change the 
tags, and will introduce any new tags if and only if the existing are not
appropriate for the users' goals. These
new tags can be organized into new XML Schemas. The HepML XML Schemas
validate a HepML block, i.e. a validator should conclude whether this set of
tags is a HepML document or not. If, we get a positive answer, an XML parser
can process the document automatically. It is essential to understand the
HepML Schemas validate the standard tags only. If a user adds new tags,
(s)he should develop a code in order to extract and use an information from
these new tags in a program. Another obvious condition for any new tags -- a
HepML block with the new tags must be an XML document. Meaning of all
standard tags is rather clear and can be derived from names of the tags (see
Appendix of the paper). The next section describes the HepML XML Schemas in
detail.

The second necessary part of HepML is an application programming interface
(API). The main goal of the API is parsing a HepML block in an event file.
Currently our programming interface is realized as a C++ library, called 
{\it libhepml}. The library consists of object classes, which correspond to
complex structures in the HepML XML Schemas, parsing classes, and serialization
classes. All
information read from a HepML block is stored in the object classes of the
API. Thus, a user should create instances of the classes in his/her program and
extract needed information from the classes. Sections~\ref{code}
and~\ref{howto} describe {\it libhepml} in detail. 

\begin{figure}[hbtp]
\begin{center}
\includegraphics[width=130mm,height=110mm]{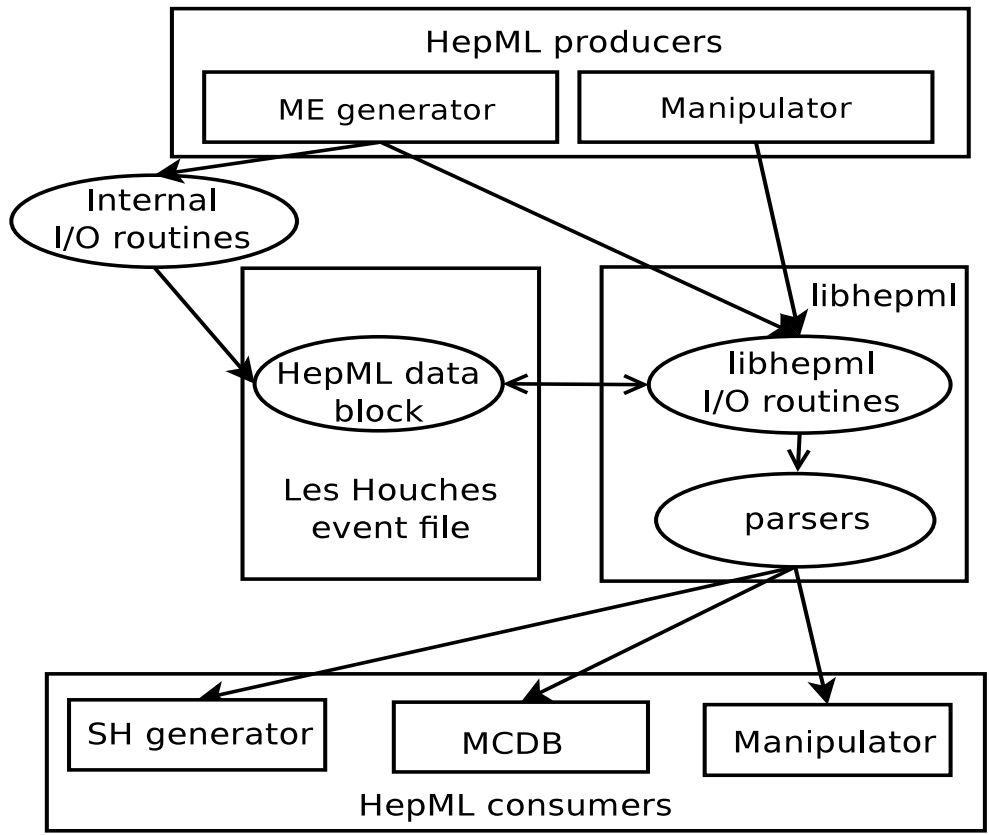}
\end{center}
\caption{
A simple scheme of interaction of programs via HepML data blocks. 
}
\label{fig:general}
\end{figure}

HepML can be useful in applications of several types. Further we follow
terminology of~\cite{Boos:2001cv}. Any ME (Matrix Element) Monte Carlo event
generator can generate a HepML block in an output event file, if the program
supports the Les Houches event format. A SH (Showering and Hadronization)
Monte Carlo event generator can use data from the block for further
processing events from the file. An information stored in a HepML block is
useful for describing event files in data bases of the files. As an example
of the DB we shall consider MCDB~\cite{Belov:2007qg}. The last class of
applications can be called event manipulators. Usually it is not a
stand-alone program, but a part of a large Monte Carlo program package or an
experimental software environment. The program processes an event file ( or
several files) and generates a new event file. For example, a manipulator can
apply a new kinematical cut and store all events passed the cut in a new
event file. Since all applied cuts are stored in HepML we should modify the
HepML block and add an information about this new cut in the output event
file. Another manipulator mixes several event files into one. In this case
we have to check whether we can mix events from the files at all. If yes, 
the program combines HepML blocks from the files in one HepML block in the 
output event file. Parsing the HepML blocks in the input event files is the
key problem in manipulators. Fig.~\ref{fig:general} represents an interaction
between programs of different types via HepML blocks and API routines needed
for that. The same manipulator can be both producer and consumer of a HepML
information.

The current HepML library depends on either Xerces~\cite{Xerces} or 
Expat~\cite{Expat}. Many developers try to exclude external dependencies as
much as possible. In fact, {\it libhepml} is a necessary tool for parsing
only. If a program writes a HepML block down into an event file, the dependency
on xml libraries can be excluded. In this case the program has to have internal
routines for generation of XML tags. Certainly, if the structure of the HepML block 
changes, the code should be modified according to the changes. 

\section{HepML XML Schemas}
\label{schema}
The HepML XML Schemas provide a general and formal description of an
information about events which should be kept in HepML files.
In other words, the Schemas represent a formal agreement how to represent 
and handle such descriptions.
Besides the Schemas can be used for automated generation of a program code 
to handle HepML documents. For example, see the description of {\it libhepml} in 
the next chapter. Though a natural way to operate with HepML documents 
is to use {\it libhepml}, having the standard and extensible XML Schemas 
allows users to make their own implementation of specific parsers, output 
routines and validators. 

Authors of MC codes can use the XML Schemas in developing of I/O routines.
If a routine is consistent with the Schemas, event files generated by
the routine can be read by another program without changes in input routines
of the program. Also the Schemas can be used for validation of event files
if the files are written according to HepML specifications. 

One of the main goals of HepML is to store all significant information 
on simulated events as well as generator input parameters and setup in 
XML documents. The first task, a detailed description of events, is done 
by the LCG group, the second task is carried out by CEDAR~\cite{Butterworth:2004mu}. 
Since these problems are complement to each other, Schemas, developed by 
both groups, are united in the main XML Schema of the HepML 
language~\textit{hepml.xsd}. Here and below we 
consider the LCG Schemas only. Lets mention the main parameters 
described in a typical HepML document. It should contain a general 
information about what kind of problem the events have been prepared for, 
a generator name, description of the physical process, description of 
the physical model in use, applied cuts, a general information on the 
event file: 
\begin{itemize}
\item 
  \textbf{General information}: title, abstract, authors, experiment and/or
  groups. 
\item 
  \textbf{Event files}: physical process/subprocesses, the number of events,
  the total cross sections and its errors, file name, location(s).
\item 
  \textbf{Physics process}: initial and final states, QCD scales, 
  parton density functions (PDFs) applied, subprocesses. 
\item 
\textbf{Used generator}: name and version, description, home page.
\item 
  \textbf{Theoretical model}: name, description, parameters and their
  values with author's description. 
\item 
  \textbf{Applied cuts\footnote{Since an event file can contain events for several independent subprocesses, 
  cuts in the HepML record are subprocess specific, i.e. there can be several sets of cuts for every subprocess 
   kept in the files.}}: cut objects, minimal and maximal possible values,
   and description. 
\end{itemize}

There are three main Schemas in the LCG part of HepML. The first XML 
Schema~{\it lha1.xsd} corresponds to the whole set of parameters composing
the first LHA agreement~\cite{Boos:2001cv}. The other two Schemas 
are~{\it sample\--description.xsd} and~{\it mcdb\--article.xsd}. 
The first one describes parameters which
are necessary to generate an XML data for an event sample. The second
Schema determines representation of data from LCG MCDB, it is used to form 
an LCG MCDB article for the sample. The CEDAR team develops other XML Schemas for
other tasks arising in the problem of automatisation of data processing in
HEP. Now all the Schemas are unified in one general formal XML Schema
\textit{hepml.xsd}. The Schema includes all the other Schemas as sub-Schemas.
This solution leaves freedom to develop new Schemas and software independently, 
but to use Schemas of both groups in one software project. All the Schemas are 
available in \cite{hemplschemas}. 

\section{Code Structure Description}
\label{code}
In this section we present a library implementing the HepML standard.
The library is called {\it libhepml}.
Classes of this library can parse HepML blocks in LHEF 
files, represent data from the blocks in computer memory, combine several 
HepML blocks in one, and serialize C++ objects into HepML documents in 
files. {\it Libhepml} consists of several types of C++ classes: 
\begin{itemize}
  \item
  {\bf object classes} represent complex types from the HepML XML Schemas.
  The main object class is ``Article''.
  This class describes a set of event files. All other object   classes are
  intended for implementation of some pieces of information in ``Article'',
  such as a physical model, a cut, beams, etc. ``Article'' includes these
  classes as internal objects.
  \item
  ``Acting'' classes. {\bf ``Parser''} is the main parsing class. All complexity 
  of parsing of XML tags and translating data to object classes is hidden in
  methods of the class. {\bf ``Mixer''} is the main class for merging HepML
  objects. {\bf ``Writer''} is the interface class for serialization of HepML
  documents.
  \item
  {\bf XML generating classes} serialize contents of an object class to 
  XML tags. 
  \item
  {\bf Parsing classes} parse HepML tags. These internal classes are 
  not intended for direct use by end-users of the library. 
  \item
  {\bf Implementing classes} translate data from parsing classes to 
  object classes. Every parsing class has its own implementing class. 
  \item
  {\bf Interacting classes} interact with classes of external libraries,
  currently with classes of Xerces or Expat.
\end{itemize}
Three first types of classes realize a user API of the library. The last 
three types of classes fulfil basic functionality in the library. 
All these basic internal classes have been prepared partly by means 
of \textit{XSD}~\cite{codesynthesis-main}
This software prepares a set of classes 
for parsing XML documents according to an XML Schema. In our case, 
the parsing and implementing classes realize the HepML XML Schemas.

Strictly speaking the ``Article'' class is a container for 
all information from a HepML block. The class interface looks like: 
\begin{verbatim} 
class Article {
 public:
    Article();
    virtual ~Article();
    int& id();
    int& id(int id);
    string& title();
    string& title(const string&);
    string& abstract();
    string& abstract(const string&);
    string& comments();
    string& comments(const string&);
    ExperimentGroup& experimentGroup();
    ExperimentGroup& experimentGroup(const ExperimentGroup&);
    vector<Author>& authors();
    vector<Author>& authors(const vector<Author>&);
    const string postDate();
    Process& process();
    Process& process(const Process&);
    Generator& generator();
    Generator& generator(const Generator&); 
    Model& model();
    Model& model(const Model&);
    CutsVector& cuts();
    CutsVector& cuts(const CutsVector&);
    vector<File>& files();
    vector<File>& files(const vector<File>&);
    vector<string>& relatedPapers();
    vector<string>& relatedPapers(const vector<string>&);
}
\end{verbatim} 
We can see the class has paired methods with the same names for 
getting/setting parameters. Information stored in the parameters 
correspond to information stored in tags of the HepML tag 
\verb|<samples></samples>| (see Appendix to the paper). For example, 
the class Generator stores an information (name, version, description, 
homepage) for a Monte Carlo generator which has been used for simulation
of events. The method ``abstract()'' returns a text of an abstract of the 
HepML document. 

The ``acting'' API classes are ``Writer'', ``Parser'', ``Mixer''. The 
first class generates HepML documents from an ``Article'' 
object. The second class is responsible for parsing one or several HepML blocks, 
e.g. ones stored in LHEF event files. It fills out an ``Article'' object.
The ``Mixer'' class merges several HepML blocks into a new block. 
``Mixer'' follows the algorithm: 
\begin{itemize}
  \item
  Sum cross sections of all event samples and combine their errors;

  \item
  Concatenate string parameters, such as model and generator descriptions, 
  authors, etc. The parameters for different HepML blocks are separated with
  semicolons;

  \item
  Compare beams. They must be the same in all HepML blocks. Otherwise 
  ``Mixer'' aborts execution and returns an error; 

  \item
  Subprocesses from all HepML blocks are combined in one array;

  \item
  Combine cut sets. A cut set is added to the array of cut sets 
  if the array does not have it yet;

  \item 
  Combining physical models is more complicated problem. If a model parameter 
  is not found in the combined model it is appended to the array of model parameters. 
  If it is found, but it has a different value, it is added only if the special 
  flag ``hepml::force::merge'' is assigned in ``Mixer::mixObjects(...)''.
  Otherwise the library aborts execution with error.
\end{itemize}

Combination of several subprocesses from several event files into one 
array of subprocesses has one subtle point. Different cuts can be 
applied in these subprocesses. Therefore, we have to unify events into 
one event sample and keep these cuts separately. In order to solve the problem we 
apply the following algorithm. Every subprocess is assigned 
an id number. This number is kept as an attribute in the tag \verb|<subprocess>|,
for example: \verb|<subprocess cutset_id="2" >|. All cuts applied for events 
of the subprocess are unified inside the tag \verb|<cutSet>| with the same 
attribute, for example: \verb|<cutSet cutset_id="2">|. All \verb|<cutSet>| are 
combined in an array and located inside the tag \verb|<cuts>|. 
Thus, the final HepML document will have two arrays of subprocesses and 
corresponding cut sets. 

The last part of the API is generating classes. They serialize C++ 
objects, corresponding to object classes, to XML code, i.e. they produce 
HepML documents. Every object class has its own generating class. 
The set of XML tags generated for the ``Article'' class can be used
both as an XML independent document or as a block of information, which 
can be included to files in other formats, e.g. event files in
the LHEF format produced with a Monte Carlo generator.

\begin{figure}[hbtp]
\begin{center}
\includegraphics[width=130mm,height=75mm]{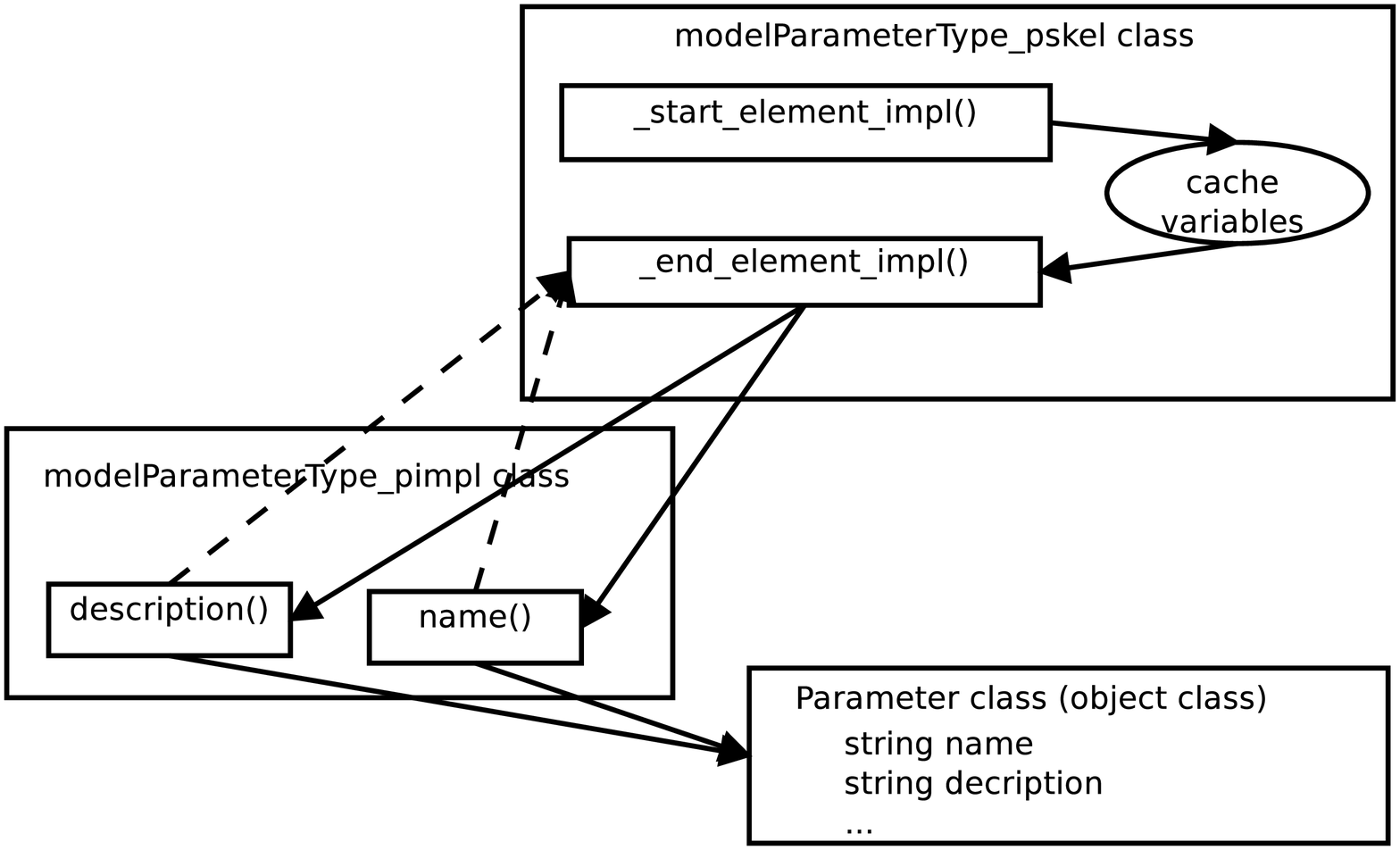}
\end{center}
\caption{
The interaction of parsing and implementing classes. Solid lines represent 
transfer of data. Dashed lines represents dependence of  ``end\_element\_impl'' on 
assigning methods. See details in text. 
}
\label{fig:parsing}
\end{figure}

Figure~\ref{fig:parsing} represents a diagram of typical interaction between 
a parsing class and an object class for filling out a HepML object. We 
use ``model parameters'' as an example. For simplicity we omit most of 
methods in the classes and keep methods needed only for the purpose of 
the picture. At fist, low-level methods of interfacing classes parse the 
whole HepML block and extract one by one tags corresponding to complex 
types in the HepML Schemas. {\it Libhepml} has a parsing class for every such 
tag. Classes of this type has a postfix ``\_pskel''.  The parsing class 
analyses the tag name and assigns the tag content to a cache variable. The 
variable can be of either a standard type (int, string, etc.) or another 
complex type (any HepML tag nested in the tag). These operations are carried 
out by the method ``start\_element\_impl''. ``start\_element\_impl'' verifies 
also whether the tag name belongs 
to the hepml namespace (a C++ realization of the vocabulary of HepML tags).
After that the method ``end\_element\_impl'' assigns the cache variable to
a corresponding internal variable in ``Article'' or another HepML class, 
via methods we call assigning methods. However, these assigning methods 
are not implemented in the parsing class. Realisation of the methods is 
the main goal of the implementing classes. Every implementing class inherits 
to a corresponding parsing class. The implementing classes are instantiated 
by the ``hepml::Parser'' class, the main ``acting'' parsing class in the API. 
Since a typical HepML block contains lots of tags, all corresponding parsing 
objects have to be combined. It happens via  callback methods called 
``parsers(...)''.
Every parsing class has its own ``parsers'' method. In terms of the HepML 
schemas the method defines tags which have to exist inside the tag, the 
parsing class corresponds to. 

All classes of {\it libhepml} can deal with tags of the HepML Schemas only. 
However, 
there can be necessity to introduce new user-defined tags in HepML blocks. If 
one needs a simple tag without any nested tags, a user can modify existing 
parsing/implementing/object classes for a complex type in the HepML Schemas. 
If a more complicated set of tags is necessary, a new object class should be 
created. There is an example of such a set of classes in the library. 

\section{How to use {\it libhepml}}
\label{howto}
{\it Libhepml} can perform tasks of three types: creating an article 
object in a program and send the object to an output stream, parse an 
HepML document in a file and keep information from the block in 
the article object, and merge several HepML documents into one. 
Below we present simplified examples for each of these tasks.

In order to use {\it libhepml}, three C++ headers should be
added to code:
\begin{verbatim}
#include <hepml/hepml.hpp>   // general object classes
#include <hepml/writer.hpp>  // to produce HepML documents
#include <hepml/parser.hpp>  // to parse and merge HepML documents
\end{verbatim}
Note: if one of two last headers are included, there is no need in the first one.

At first an instance of the main object class ``Article'' should be 
created\footnote{We assume the hepml namespace is defined in the source 
file via the standard C++ instruction ``using namespace hepml;''. So, all 
{\it libhepml} classes are used without the namespace prefix ``hepml::''}:
\begin{verbatim} 
  Article a;
\end{verbatim}
Lets fill out the object with information. There is two different ways how
to assign values to parameters of the object. For example: 
\begin{verbatim} 
  a.title("p,p->Wbbj->l,nu,b,b,j process from CompHEP");
\end{verbatim}
or 
\begin{verbatim} 
  a.abstract() = "There are about 1.1M events for the process 
                  p,p->W,b,b,j with leptonic decays of W-boson.";
\end{verbatim}
The object classes have several arrays for authors, model parameters, subprocesses, 
and cuts. Lets create an author record and add it to the article object:
\begin{verbatim}
 Author author;
 author.firstName() = "James";
 author.lastName() = "Johnson";
 author.email() = "James.Johnson@nospamcern.ch";
 author.experiment() = "CMS";
 author.experimentGroup() = "CMS Top group";
 author.organization() = "CERN";
 a.authors().push_back (author);
\end{verbatim}
After that we should introduce and describe a physical process in the
article. At first, we describe initial beams: 
\begin{verbatim} 
  Process p;
  p.beam1().particle = Beam::Particle("p");
  p.beam1().energy = Beam::Energy(7000, Beam::Energy::GeV);
  p.beam1().pdf.name = "CTEQ6L";
  p.beam1().pdf.lhaPdfSet = 3;
  qc = QcdCoupling();
  qc.lambda = 2.0;
  qc.nLoopsAlpha = 2;
  qc.nFlavours = 5;
  p.beam1().qcdCoupling = qc;
\end{verbatim}
The second beam is set in the same manner (certainly, beam2() should be used
instead of beam1()). We have to describe the final state, assign information 
on a cross section, and add a description of the process to our article: 
\begin{verbatim}
  p.finalState() = "l,nu,b,b,j";
  p.finalStateNotation().plain = "l,nu,b,b,j";
  p.finalStateNotation().html = "<i>l,nu,b,b,j</i>";
  CrossSection cs = CrossSection( 22.78, CrossSection::pb );
  cs.errorPlus = cs.errorMinus = 0.02;
  p.crossSection() = cs;
  a.process() = p;
\end{verbatim}
If we have several subprocesses in the process, we add them to the subprocesses
array in the article object:
\begin{verbatim}
  Subprocess sp;
  sp.notation() = "u,D -> nm,M,G,b,B";
  sp.crossSection().value = 1.3221;
  sp.crossSection().unit  = CrossSection::pb;
  sp.crossSection().errorPlus = 
  sp.crossSection().errorMinus = 2.68e-03;
  a.process().subprocesses().push_back( sp );
\end{verbatim}
We should keep information on a generator and append the information to 
the article:
\begin{verbatim}
  gen.name() = "CompHEP";
  gen.version() = "4.5.2";
  gen.homepage() = "http://comphep.sinp.msu.ru";
  gen.description() = "Funny Monte Carlo event generator";
  a.generator() = gen;
\end{verbatim}
The next important step is an introduction of a physical model:
\begin{verbatim}
  Model& m;
  m.name() = "Standard Model";
  m.description() = "There can be a long and detailed 
                     description of the model";
\end{verbatim}
Model parameters should be added one by one:
\begin{verbatim}
  Model::Parameter param = Model::Parameter( "Ms", "0.117");
  param.mathNotation().plain = "Ms";
  param.mathNotation().html  = "m<sub>s</sub>";
  param.description() = "parameter1 description";
  m.parameters().push_back(param);
\end{verbatim}
If any kinematical cuts have been applied we can add description of the
cuts. 
\begin{verbatim}
  Cut cut;
  cut.object() = "M(l,nu)";
  cut.minValue() = "100 GeV";
  cut.maxValue() = "";
  cut.objectNotation().html = "html object notation";
  a.cuts().push_back(cut);
\end{verbatim}
After that we assign an information about the event file:
\begin{verbatim}
  File f;
  f.eventsNumber() = 195644;
  f.size() = 407736817;
  f.crossSection() = CrossSection(22.78, CrossSection::pb);
  f.checksum().type = File::Checksum::mda5;
  f.checksum().value = "8957a237dc062b96987a21c86774eb5e";
  a.files().push_back(f);
\end{verbatim}

As soon as the article is created and filled out with necessary
information, it can be serialized to an output stream, for example to the 
standard output stream: 
\begin{verbatim}
  hepml::Writer writer;
  std::cout << writer.toHepml( a );
\end{verbatim}
In this case the article will be printed out in the form of a HepML block.
If a parameter in the ``Article'' object is necessary (according to the HepML
Schemas) and it is undefined, it will be assigned to a reasonable value,
an empty string for a {\it string} parameter, 0 for {\it int}, etc. If the
parameter default value is specified in the HepML Schemas, the value will
be used. 

The second typical task performed by {\it libhepml} is HepML parsing. In order
to parse a HepML block in a file and fill out an instance of the ``article''
class we should define instances of the ``Article'' and ``Parser''
classes and parse the LHEF files by means of the parseObject object:
\begin{verbatim}
  Article a;
  Parser parser;
  ::std::string file = "hepml_examples/general/example1.xml";
  parser.parseObject (a, file);
\end{verbatim}
After that, all information from the HepML block is available in the object
``a''. We can manipulate the information. For example: 
\begin{verbatim}
  cout << "Generator name: " << a.generator().name() << endl;
  cout << "Model: << a.model().name() << endl;
  cout << "  with parameters:"
  for (int i(0); i < a.model().parameters().size (); ++i) {
    cout << "    name: " << a.model().parameters()[i].name() <<
      ", value: " << a.model().parameters()[i].value() << endl;
  }
\end{verbatim}

If we have several LHEF files and want to merge HepML blocks in the files, 
we can use the ``Mixer'' class and the mixObjects method; the second 
argument should be a vector of file names. 
\begin{verbatim}
  Article article;
  Mixer mixer;
  vector< ::std::string> files(2);
  files[0] = "hepml_examples/general/example1.xml";
  files[1] = "hepml_examples/general/example2.xml";
  try {
    mixer.mixObjects(article, files);
  }
  catch (const ::std::exception m) {
        m.what();
        return 1;}
\end{verbatim}
The merging algorithm was explained in detail in the previous section. The only 
subtle point here is merge of models with a parameter, which has different values 
in different HepML blocks. In order to merge the HepML blocks in this case 
we should use a special flag:
\begin{verbatim}
  ...
  try {
    mixer.mixObjects(article, files, force::merge);
  }
  catch (const ::std::exception m) {
        m.what();       
        return 1;}
\end{verbatim}

\section{Current HepML applications}
\label{application}
\subsection{LCG MCDB}
In the last years there was a need for common place to store sophisticated 
Monte Carlo event samples prepared by experienced theorists. Also such 
samples should be accessible in a standardized manner to be easily imported 
and used in experiments' software environments. The main motivation behind 
the LCG MCDB project~\cite{Belov:2007qg,lcg_mcdb_url} is to make the sophisticated 
Monte Carlo event samples and their structured descriptions available for 
various groups of physicists working at the LHC. All these data from MCDB 
are accessible for end-users in several convenient ways from Grid, on the 
Web, and via an application program interface. 

The main content of MCDB are event files and their detailed descriptions. 
These descriptions are fully compatible with the information which can be 
provided in HepML blocks. So, the main way to automate an access to MCDB is 
to use HepML documents in interaction with MCDB. An event sample description 
can be both exported from MCDB or uploaded to the data base. In other words, 
an MCDB article can be obtained as a HepML document. Otherwise, a new article 
can be created automatically in MCDB using a HepML description of an event file. 

\begin{figure}[hbtp]
\begin{center}
\includegraphics[width=140mm,height=80mm]{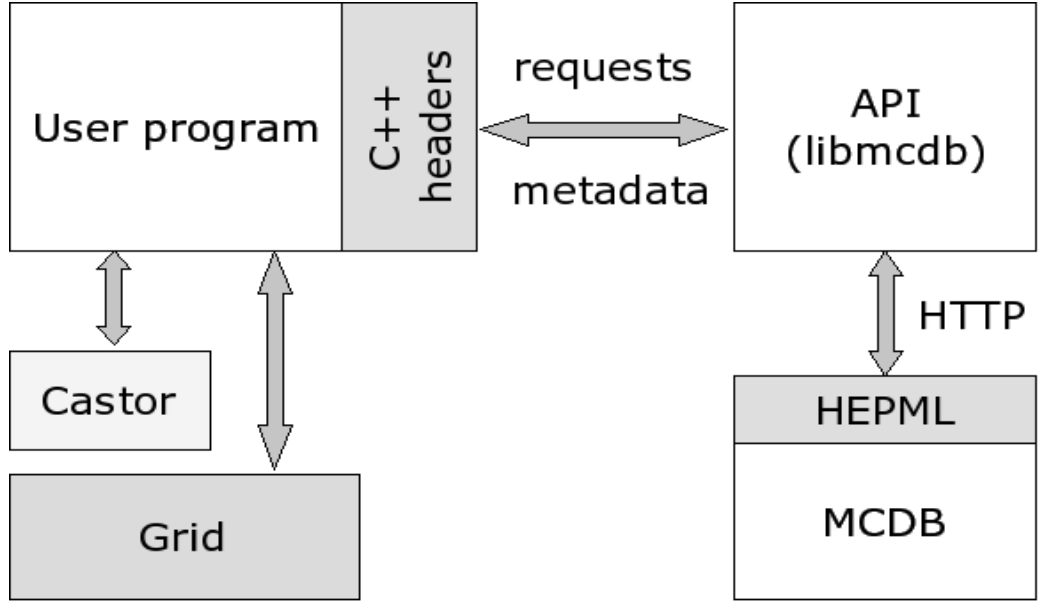}
\end{center}
\caption{
A simple scheme of interaction of MCDB and CMSSW in Monte Carlo 
productions. 
}
\end{figure}

The CMS collaboration~\cite{cms} already uses MCDB in its productions of 
simulated events. Event files can be downloaded from MCDB by means of internal 
routines of the CMSSW software simulation environment. The routines are 
based on classes of {\it libhepml}. There is also an option to upload new event 
files to MCDB using a special uploading script. The script supports LHEF 
files with two types of header blocks, HepML and MadGraph. 


\subsection{CompHEP Monte-Carlo generator}

The Monte Carlo event generator CompHEP produces events on the partonic 
level for particle decays and particle collisions at colliders. The main 
file format for the event files in CompHEP is LHEF, although 
two obsolete native event file formats are still in use for backward 
compatibility. More information about the program can be found 
in~\cite{Boos:2004kh, Boos:2009un}.

Since CompHEP generates partonic level events it is a natural target for 
implementing HepML. Currently, by default, CompHEP needs neither an xml 
library no {\it libhepml} in order to produce a HepML block, since the 
task of generating XML code is rather simple programming problem, and the 
output format is rather stable. The block is constructed with internal 
CompHEP routines. However, CompHEP can be compiled with the libxml2 
library. It is a standard GNOME XML library. 

There are several problems where CompHEP HepML block should be parsed and 
modified. It means the libxml2 parser should be used. CompHEP generates events 
per subprocess, i.e. a physical process with fundamental model particles in 
the initial and final states. However, in many cases we need to sum over 
several initial or final states. For example, this happens if we have a 
composite particle, like proton, in the initial state. Therefore we come up 
with several event files, which should be combined into one event sample. 
This problem is a particular case of the merging problem we discussed in 
Sect.~\ref{code}. HepML blocks of subprocess event files should be merged 
into one block. There is a special mixing program, called ``mix'', for that 
in CompHEP. If ``mix'' is linked against libxml2 it combines all 
HepML blocks from input files and adds a new block to the final event file. 

Since two authors of the paper are also members of the CompHEP collaboration, 
we plan to expand usage of HepML in CompHEP. Namely, we plan to add 
support of HepML to the ``addcut'' and ``cascade'' programs. The 
former program applies a new cut to events kept in a CompHEP event file. 
So we have to add the cut to the HepML block in the output event file. 
The letter program replaces heavy decaying resonances in final states 
with their decay products. Therefore the output event file contains 
subprocesses with different final states. A list of these new subprocesses 
should replace the old list of subprocesses. In the future, we plan to 
use {\it libhepml} instead of libxml2 in CompHEP, since this will help 
significantly to simplify codes of programs, where HepML is used. 

\section{Conclusion and Plans}
\label{conclusions}
In the article we present HepML, a new markup language for describing events 
on the partonic level in a uniform and flexible manner. Blocks of HepML tags 
can be painlessly implemented into LHEF files. HepML allows Monte Carlo event 
generators to prepare self-documented event files. It means we have all 
necessary information about events, such as physical model, applied cuts, 
etc., inside of the event file. HepML is constructed as an extensible language. 
Users can extend the set of XML schemes or add new tags into the format. 
The only requirement to these new tags is that they should follow XML rules. 
The structure of a HepML block is defined with several HepML XML Schemas. 

HepML is equipped with an API in the form of a C++ library. It is called 
{\it libhepml}. The library consists of object classes for representing 
information from a HepML block in computer memory, parsing classes, and 
classes for serialization the object classes into HepML tags. This library 
can use either Xerces or Expat XML parser for low-level parsing of XML tags. 
There is a possibility to add processing of new user-defined tags into 
the library. {\it Libhepml} provides a unified interface for the automatic 
event description at different levels of Monte-Carlo simulation in HEP. 

The developed HepML schemes, documentation and code of the {\it libhepml} library 
are available publicly on the MCDB web server~\cite{lcg_mcdb_url}. 

There are several projects which have already started to exploit HepML. LCG 
MCDB uses HepML information in all its external interfaces. The CompHEP Monte 
Carlo event generator adds HepML blocks 
into event files and uses the blocks for mixing of several event files into 
one event file. HepML has been implemented in the software environment of 
the CMS collaboration (CMSSW) in order to document externally simulated 
event samples, kept in LCG MCDB. 

We are going to develop the project further in the framework of an open
source project~\cite{hepforge} and encourage people interested in 
development of XML-based formats to join the project. 

\section{Acknowledgements}
This work was supported by the RFBR under grant 07-07-00365-a
and the FASI state contract 02.740.11.0244. 
Participation of A.~S. in the project was partly supported by the UK Science 
and Technology Facilities Council. We also acknowledge the LCG collaboration 
for support and hospitality at CERN. 

\section{Appendix. HepML tags.}
\label{appendix}
The main LCG HepML Schema consists of a number of tags. 
Here we collect information about the tags:

\begin{itemize}
\item
  \verb|<samples>| is the root tag of a HepML block. It contains information 
  on each event file (in the \verb|<files>| tag) and a common description of 
  events in the files (in the \verb|<description>| tag).
\item
  \verb|<description>| describes an article. It should have several simple 
  tags\footnote{A simple tag contains a text only and does not have any nested tags.}
  (\verb|<title>|, \verb|<abstract>|, \verb|<authorComments>|, \verb|<relatedPapers>|) 
  and several tags with nested tags (\verb|<experimentGroup>|, \verb|<generator>|, 
  \verb|<model>|, \verb|<process>|, \verb|<cuts>|, \verb|<authors>|).
\item
  \verb|<authors>| contains a list of all authors of the event sample. 
\item
  \verb|<author>| describes an author, i.e. contains tags for first 
  and last author's names, his/her email and affiliation (group, 
  experiment, organization). 
\item
  \verb|<title>| contains a title of the article. 
\item
  \verb|<abstract>| contains an abstract of the article. 
\item
  \verb|<relatedPapers>| contains a list of related articles. 
\item
  \verb|<authorComments>| contains additional authors' comments to the article. 
  All information not marked up with tags should be stored within the tag. 
\item
  \verb|<experimentGroup>| contains an information about an experiment and/or 
  a group, which produce the events. It consists of several simple tags 
  (\verb|<experiment>|, \verb|<group>|, \verb|<responsiblePerson>|, and 
  \verb|<description>|). 
\item
  \verb|<generator>| contains an information on a Monte Carlo event generator 
  used. There are several simple tags  in the tag: \verb|<name>|, \verb|<version>|, 
  \verb|<homepage>|, and \verb|<description>|. 
\item
  \verb|<model>| describes the physical model for the events, i.e. all 
  parameters in the model. It has \verb|<name>| and \verb|<description>| 
  tags, and an array of parameters within \verb|<parameters>| tag. Each 
  \verb|<parameter>| describes an element of the model by means of four 
  tags: \verb|<name>|, \verb|<value>|, \verb|<description>|, and \verb|<notation>|.
\item
  \verb|<process>| describes a physical process. It contains several tags: 
  \begin{itemize}
  \item
     \verb|<beam1>,<beam2>| describe initial beams. It means each beam tag 
     defines the particle info (tag \verb|<particle>|), energy (tag \verb|<energy>|), 
     structure functions (tag \verb|<pdf>|), and an information on the QCD coupling 
     related to the structure function (tag  \verb|<QCDCoupling>|). 
  \item
     \verb|<QCDCoupling>| defines an information on the QCD coupling if it is not defined 
     in beams.
  \item
     \verb|<finalState>| defines a final state for the process.
  \item
     \verb|<crossSection>| sets the total cross section of the process. 
  \item
     \verb|<subprocesses>| defines a list of subprocesses for this physical 
     process. Each subprocess has the following characteristics: 
     notations for initial and final states, the total cross section, and 
     factorization and renormalization scales.
  \end{itemize}
\item
  \verb|<cuts>| contains a list of applied cuts. These cuts are grouped in 
  several cut sets (tag \verb|<cutSet>|). \verb|<cutSet>| has a special 
  attribute (\verb|cutset_id|), which couples a \verb|<cutSet>| tag and a 
  \verb|<subprocess>| tag. It means if the tags have the same value 
  of the attribute, cuts from the \verb|<cutSet>| tag are applied for events 
  of the subprocess. 
\item
   \verb|<cut>| defines a cut. 
\item
  \verb|<files>| is a list of event files. This element can be both used separately 
  and inside the \verb|<samples>| tag.
\item
  \verb|<file>| contains an information about one event file. A type of the 
  file can be specified in the attribute \verb|type|. It has the following 
  nested tags: 
  \begin{itemize}
  \item
     \verb|<eventsNumber>| -- the number of events in the file.
  \item
    \verb|<crossSection>| -- the total cross section for events in the file.
  \item
    \verb|<fileSize>| -- the file size in bytes.
  \item
    \verb|<checksum>| -- a check sum of the file (a type of the check sum is specified in the \verb|type| attribute).
  \item
    \verb|<comments>| -- an author's comment for the file.
  \item
    \verb|<location>| -- a list of locations of the file (CASTOR, Grid or others).
  \end{itemize}
\end{itemize}


\end{document}